\documentclass[12pt,draft]{article}
\title{Galileo's Discovery of Scaling Laws}
\author{Mark A. Peterson \\ Physics Department \\ Mount Holyoke
College, South Hadley MA 01075}
\begin{document}
\maketitle
\begin{abstract}
Galileo's realization that nature is not scale invariant,
motivating his subsequent discovery of scaling laws, is traced to
two lectures he gave on the geography of Dante's Inferno.
\end{abstract}
\section{Introduction}

Galileo's last and crowning achievement was the {\bf Two New
Sciences} \cite{TNS}, a dialogue in four days.  The second half of
this book, the third and fourth days, describes his solution to
the longstanding problem of projectile motion, a result of obvious
importance and the birth of physics as we know it.  But this was
only the second of his two new sciences.  What was the first one?

{\bf Two New Sciences} begins in the Venetian Arsenal with a
discussion of the effect of scaling up or scaling down in
practical construction projects, like shipbuilding.  The
conversation veers away into a dazzling variety of topics, but on
the second day, after that long digression, it returns to a
serious analysis of scaling, especially in the context of the
strength of materials. According to the publisher's foreword, it
is this topic that should be understood as the first of the two
new sciences \cite{TNS2}. Since it was the publisher and not
Galileo who gave the book its title, this must be what the title
means. It should be noted that Galileo was unhappy with the title,
perhaps because it seems irrelevant to much of the ingenious
speculation of the First Day \cite{TNS1}. It is clear, though,
that Galileo did assign enormous importance to the problem of the
strength of materials, an importance that modern readers have
found more than a little puzzling.

I will show that the key to much of what is strange in {\bf Two
New Sciences} is to be found in two rather neglected early
lectures given by Galileo on the shape, location, and size of
Dante's Inferno \cite{lecture}.  The text of these lectures is
readily available in the standard 20-volume {\bf Opere} of Galileo
among the ``literary" writings in Volume 9.  My reconstruction of
its actual significance, which is not at all literary, is the
subject of this paper.

\section{Scaling in the Two New Sciences}

{\bf Two New Sciences} begins with the subject of scaling.
Galileo's observations on scaling in general are ingenious and
elegant, and entirely deserving of the prominent place he gives
them. These ideas are basic in physics, and are introduced, in
some fashion, in Chapter 1 of most introductory physics texts
under the heading of dimensional analysis. We could even say that
modern renormalization group methods are just our most recent way
to deal with problems of scale, still recognizably in a tradition
pioneered by Galileo. It is true that Galileo didn't have the
algebraic notation to do dimensional analysis the way we do, but
his insights within a restricted arena are the same as ours.

An example of such an insight is ``the surface of a small solid is
comparatively greater than that of a large one" \cite{TNS3}
because the surface goes like the square of a linear dimension,
but the volume goes like the cube. Thus as one scales down
macroscopic objects, forces on their surfaces like viscous drag
become relatively more important, and bulk forces like weight
become relatively less important.  Galileo uses this idea on the
First Day of the {Two New Sciences} in the context of resistance
in free fall, as an explanation for why similar objects of
different weight do not fall exactly together, but the lighter one
lags behind.

Even though the idea is completely general, most of the discussion
on scaling is directed specifically to the strength of materials.
That subject is introduced immediately in {\bf Two New Sciences}
with the assertion that large ships out of water risk breaking
under their own weight, something that is not a concern with small
ships \cite{TNS4}.  The same subject occupies most of the Second
Day, devoted to the strength of beams. The question is put in an
especially paradoxical way by Sagredo, who says ``Now, since
mechanics has its foundation in geometry, where mere size cuts no
figure, I do not see that the properties of circles, triangles,
cylinders, cones and other solid figures will change with their
size.  If therefore a large machine be constructed in such a way
that its parts bear to one another the same ratio as in a smaller
one, and if the smaller is sufficiently strong for the purpose for
which it was designed, I do not see why the larger also should not
be [sufficiently strong] ..." \cite{TNS4}.  This is a remarkable
argument.  It combines (1) faith that physics ``is" geometry, and
(2) the scale invariance of the theorems of geometry. For the
purpose of this paper we will call it Sagredo's theory of scale
invariance \cite{conformal}.

{\bf Two New Sciences} modifies Sagredo's theory of scale
invariance.  Specifically, Galileo adds material properties to (1)
above, while still preserving the essential ``geometrical" nature
of physics. As Salviati puts it much later, on the Second Day,
``... these forces, resistances, moments, figures, etc. may be
considered in the abstract, dissociated from matter, or in the
concrete, associated with matter. Hence the properties which
belong to figures that are merely geometrical and non-material
must be modified when we fill these figures with matter and
therefore give them weight." \cite{TNS5} These modifications break
scale invariance, but they do not change the essence of the
theory, which is mathematical: ``Since I assume matter to be
unchangeable and always the same, it is clear that we are no less
able to treat this constant and invariable property in a rigid
manner than if it belonged to simple and pure
mathematics."\cite{TNS20}

A typical example of the new considerations that arise when ``we
fill these figures with matter" starts with the breaking of a
wooden beam.  If the beam breaks, all its longitudinal fibers
break, and since their number is proportional to the
cross-sectional area of the beam, the beam's strength in
withstanding a longitudinal pull should be proportional to its
cross-sectional area.  If a supported beam is subject to a
transverse force, however, the associated torque must be balanced
by the first moment of the stress in the fibers, which introduces
an additional factor of the diameter of the beam \cite{TNS7}. Thus
the beam's strength in withstanding a transverse force should be
proportional to its diameter cubed \cite{secondmoment}.  If the
beam is made longer, the torque due to the same transverse force
applied at a geometrically similar position is proportionately
greater and the beam is proportionately weaker, and so forth.
There follow Galileo's very famous observations on why animals
cannot be simply scaled up, but rather their bones must become
proportionately thicker as they get larger \cite{TNS6}.

All of this material is developed in a series of eight
propositions with geometrical proofs. It is clearly the
geometrical framework which makes these scaling laws a ``science."
After the eight propositions there is some discussion and then a
return to geometry, considering certain more detailed questions,
such as the problem of locating the weakest point in a beam, i.e.,
where it would break, and the problem of finding a beam which
would have no unique weakest point, but would be, in a sense,
equally strong everywhere.  These last topics are clearly later
than the original scaling theory of the eight propositions.

A persistent oddity on the Second Day is a continuing
preoccupation with beams that break under their own weight, the
same phenomenon that began the whole discussion in the Venetian
Arsenal, except that there it was ships that break under their own
weight.  Another oddity is the sudden change of subject on the
First Day. The conversation, which has just barely gotten
underway, is about the strength of ropes, made of fibers, but the
question of non-fibrous materials is raised, such as metal or
stone. Simplicio wants to know what gives such materials their
strength. Just before this, and perhaps intended to motivate it,
we had heard about a marble column that broke under its own weight
\cite{TNS9}. Although Salviati hesitates to take time over this,
Sagredo says ``But if, by digressions, we can reach new truth,
what harm is there in making one now...?" \cite{TNS8} and the
conversation goes off for the rest of the day into speculations
about the atomic theory of matter, cohesive forces among atoms,
and questions of the infinitesimally small and infinitely large,
among other things. The discussion is diffuse but occasionally
brilliant, even if nothing definitive can be said about what holds
matter together. The opening lines of the Second Day confirm,
however, that the central purpose of all this was that the scaling
theory should apply to all solid materials and not just to fibrous
materials like wood. Taking everything together, there is an odd
hint of a problem in the background which motivates all the
discussions of the first two days but which is never actually
stated, some object of stone or metal that breaks under its own
weight when it is scaled up. This is what the conversation seems
to be circling around, without ever quite saying so.

In seeking to understand what the {\bf Two New Sciences} is really
about, we should recall how it came to be written. Galileo was
remarkably reticent about publishing: before the {\bf Starry
Messenger} of 1610, the year he turned 46, he had published hardly
a thing.  His astronomical discoveries of that year brought him
instant fame, and he parlayed this into his appointment as
Mathematician and Philosopher to the Grand Duke of Tuscany.  In
the negotiations leading up to that appointment he described the
many books he wanted leisure to write, based on the research he
had been quietly doing at Padua, but these books did not appear.
His life at the Tuscan court was fraught with controversy, and the
books he actually wrote were reactions to events and new
discoveries, the {\bf Discourse on Bodies in Water}, and the {\bf
Sunspot Letters}. When the Holy Office declared the opinion of
Copernicus formally heretical in 1616, his fondest project was
closed off to him.  He silenced himself for several years, and
when he finally wrote again ({\bf The Assayer}), after much
pressuring from his friends, it was part of a testy controversy
that made him new enemies. The debacle over his great {\bf
Dialogue Concerning the Two Principal World Systems} and the
disastrous trial of 1633 should have been the end of his career.
Yet somehow he found the strength to begin writing what would
become the {\bf Two New Sciences}. And from somewhere in his past
research he drew the scaling theory, and placed it first.

Just where the scaling theory comes from is a bit mysterious,
since there are few surviving references to it in the record.  It
was certainly complete by 1612, however, the year Galileo
published the {\bf Discourse on Bodies in Water}.  The most vexing
problem Galileo deals with there is the problem of a thin lamina
of material denser than water which nonetheless floats on the
surface, supported by surface tension, as we would say now.  At
the very end of the Discourse, Galileo points out that if the
lamina were supported at its perimeter, a thin sample (say of
square shape and fixed thickness) would be more easily supported
if it were cut into many smaller squares, because in reducing the
size, the area of each piece, and hence its weight, would go down
faster than the perimeter, which is the source of its support.
Galileo presents this argument purely hypothetically, as contrary
to fact, since his own idea of what supports the lamina is quite
different. Peculiar details aside, it suffices for our purpose
that this is an explicit use of the scaling theory as early as
1612.

An even earlier reference, although not as explicit, occurs in a
letter from Galileo to Antonio de'Medici in 1609 \cite{Antonio}.
This remarkable letter is in effect a one page outline of the {\bf
Two New Sciences}! Galileo is describing, he says, his most recent
investigations, saying ``And just lately I have succeeded in
finding all the conclusions, with their proofs, pertaining to
forces and resistances of pieces of wood of various lengths,
sizes, and shapes, and by how much they would be weaker in the
middle than at the ends, and how much more weight they can sustain
if the weight were distributed over the whole rather than
concentrated at one place, and what shape wood should have in
order to be equally strong everywhere:  which science is very
necessary in making machines and all kinds of buildings, and which
has never been treated before by anyone."  He then goes on to
describe certain insights into the motion of projectiles.  This
letter makes it clear that the basic organization of {\bf Two New
Sciences} was already in his mind in 1609, and even seems to say
that the results were at that time quite fresh.  On closer
inspection, though, one sees that the detailed results described
here belong to the last section of the Second Day, building on the
basic results of the first eight propositions, which must
therefore be even earlier material.  That the projectile theory
was still incomplete at this point, and also comes later in {\bf
Two New Sciences}, suggests (what is plausible in any case) that
the order of treatment there is chronological. That is, the
scaling theory comes first in {\bf Two New Sciences} because it
literally was first, and we will have to look earlier to find it.

The next suggestion for the origin of the scaling theory might
well come from {\bf Two New Sciences} itself.  The opening scene
in the Venetian Arsenal is peculiarly effective -- why not take it
at face value?  Galileo was obviously familiar with the Arsenal,
and it is known that he was consulted there on ship design.  What
likelier place could there be to encounter problems of scale, in
just the way that he says?  This could place the origin of the
scaling theory as early as 1592, the year he became Professor of
Mathematics at Padua.  On closer reading, though, one sees that
even if Sagredo is surprised by the failure of scale invariance in
the shipyard, Salviati is not.  He already understands it, and is
immediately ready to explain it all to Sagredo.  What the scene
really says is that Salviati's mind (read Galileo's) was already
prepared to understand what he saw in the Arsenal, because he had
already done the scaling theory.  This means we must search for
the scaling theory still earlier, at the very beginning of
Galileo's career, or perhaps, in a sense, even earlier than that.

\section{The Lectures on Dante's Inferno}

Galileo enrolled at the University of Pisa to study medicine at
the age of 17, and dropped out at the age of 21.  He spent the
next several years studying mathematics independently, especially
Euclid and Archimedes, and he did tutoring in mathematics. While
living at home he assisted his father in remarkable experiments on
the pitch of plucked strings under known tension \cite{drake}.
Fifty years later he briefly summarized their experimental results
in {\bf Two New Sciences}, the First Day \cite{TNS10}.  He proved
some theorems on centers of gravity, in the style of Archimedes,
and he also included those in {\bf Two New Sciences}, as an
appendix. It is clear that {\bf Two New Sciences} contains some
very early material.

The young Galileo hoped to make a reputation in mathematics with
his theorems, and he sent them to a number of Italian
mathematicians.  He was fortunate to get a favorable reply from
Guidobaldo del Monte, Inspector of Fortifications to the Grand
Duke of Tuscany, someone in a position to help him \cite{drake2}.
They corresponded.  When the chair of mathematics at Pisa became
open, someone, probably Guidobaldo but perhaps his even more
illustrious brother Francesco, who had just been made a cardinal
\cite{biagioli}, arranged for Galileo to be invited to address the
Florentine Academy to give two lectures on mathematical topics
\cite{lecture}.  It was in effect a seminar talk and job
interview.  Galileo's lectures charmed his audience. Within a few
months the young dropout was Professor of Mathematics at Pisa.

Galileo's audience at the Florentine Academy was not a
mathematical one.  The Florentine Academy was a creation of the
Medici dynasty (which had ascended to the nobility only in the
immediately previous generation), and had as one of its chief
functions the glorification of the Medici in every intellectual
arena \cite{Biagioli}.  It was far more important for Galileo to
play to this predilection of his audience than to display
mathematical erudition. In the event he brilliantly combined a
clear exposition of mathematics with a topic Florence loved to
hear, their great poet Dante, and in particular the geometry of
Dante's Inferno, based on evidence from the poem.  Although
Galileo did introduce certain original material in his lecture, he
did not call too much attention to this, and represented himself
rather as describing two previous rival attempts to determine the
plan of Hell.  One of these was by Antonio Manetti, who is
represented as a member now deceased of the Florentine Academy
itself \cite{Manetti}. The other plan, a later attempt, was by
Alessandro Vellutello, not a Florentine \cite{Vellutello}. Galileo
begins in a way that seems evenhanded, but as the second lecture
proceeds, on Vellutello's plan, he becomes more and more sarcastic
until in the end he seems to be defending the virtuous Florentine
Manetti against the ridicule of the stupid and thoughtless
Vellutello, to the delight of his audience, no doubt. This was how
a Medici intellectual should defend the honor of Florence!

Only the large scale features of these plans need concern us here.
Manetti's Inferno is a cone-shaped region in the earth, with its
vertex at the center and its base on the surface, centered on
Jerusalem.  But since Galileo is a master of exposition, let him
describe it:  ``... imagine a straight line which comes from the
center of the earth (which is also the center of heaviness and of
the Universe) to Jerusalem, and an arc which extends from
Jerusalem over the surface of the water and the earth together to
a twelfth part of its greatest circumference: such an arc will
terminate with one of its extremities on Jerusalem; from the other
let a second straight line be drawn to the center of the earth,
and we will have a sector of a circle, contained by the two lines
which come from the center and the said arc; let us imagine, then,
that the line which joins Jerusalem to the center staying fixed,
the other line and the arc should be moved in a circle, and that
in such motion it should go cutting the earth, and move itself
until it returns to where it started. There will be cut from the
earth a part like a cone; which, if we imagine it to be taken out
of the earth, there will remain, in the place where it was, a hole
in the form of a conical surface, and this is the Inferno."  As an
aside to the mathematically adept, Galileo gave the volume of this
region, which he knew from his study of Archimedes.

The various levels of Manetti's Inferno are regularly spaced, for
the most part, with 1/8 the radius of the earth between each level
and the next.  In particular the first level, Limbo, is at a depth
of 1/8 the radius of the earth below the surface, and the shell of
material down to this depth forms a cap of this thickness over the
whole of Hell. Vellutello's Inferno, by contrast, is much smaller,
located near the center of the earth, and only about 1/10 the
radius of the earth in height, making it, as Galileo is quick to
say, ridiculously small, only 1/1000 the volume of Manetti's.

Near the end of his presentation, Galileo says this: ``Here one
might oppose that the Inferno cannot be so large as Manetti makes
it, since, as some have suspected, it doesn't seem possible that
the vault that covers the Inferno could support itself and not
fall into the hole, being so thin, as is necessary if the Inferno
comes up so high. And especially, beyond being no thicker than the
eighth part of the radius of the earth, which is 405 miles more or
less, some of it must be removed for the space of the Grotto of
the Uncommitted, and even more must be removed [on the top] for
the very great depth of the sea. To this one answers easily that
such a thickness is more than sufficient; for taking a little
vault which will have an arch of 30 braccias, it will need a
thickness of about 4 braccias, which not only is enough, but even
if you used just 1 braccia to make an arch of 30 braccias, and
perhaps just 1/2, and not 4, it would be enough to support itself;
and knowing that the depth of the sea is a very few miles, or
better, even less than one mile, if we believe the most expert
sailors, and assigning as many miles as seem necessary for the
Grotto of the Uncommitted, a determinate measure not being given
by the Poet, if this together with the depth of the sea comes to
100 miles, the said vault will still be very thick, and far more
than is necessary to hold itself up."  Since in Galileo's units
the earth's radius is about 3200 miles, and 1/8 of that is 400
miles, it is clear that he is describing a scale model of the roof
of the Inferno, including a certain anteroom hollowed out of it,
at a scale of about 1 braccia to 100 miles.  A normal man is 3
braccias tall, so the model suggests a large domed roof, somewhat
smaller than the famous Brunelleschi dome of the Florentine
cathedral which, as Galileo says, is less than 4 braccias thick
and supports itself beautifully.  This is a convincing argument
that Manetti's model can support itself -- but only until you
realize that the argument assumes scale invariance! Could you
really scale it up by a factor of 100,000?  Absolutely not!  The
scaled up version is effectively weaker by that enormous factor
and would immediately collapse of its own weight.

\section{Discussion}
When Galileo realized his mistake, probably just a short time
later, it must have struck him like a lightning bolt. This is just
how Sagredo reacts, in fact, in strangely emotional language, at
the beginning of {\bf Two New Sciences} where Salviati asserts
that nature is {\em not} scale invariant: ``My brain ... reels. My
mind, like a cloud momentarily illuminated by a lightning-flash,
is for an instant filled with an unusual light, which now beckons
to me and which now suddenly mingles and obscures strange, crude
ideas..." \cite{TNS11}

We need look no further to know why the problem of scaling and the
strength of materials had urgent meaning for Galileo.  There is
nothing hypothetical about this.  It is plain in the record, with
no ambiguity at all.  He had made a gigantic blunder in the
Inferno lectures, sufficient to turn his whole argument on its
head, and with it his claim to be an intellectual champion of his
country and his sovereign, on whom his young career depended.

It may be difficult, from a modern point of view, to take
Sagredo's overwrought words seriously. Yet the situation was
serious enough, apparently, to force Galileo to concentrate on the
scaling problem and develop a new theory almost immediately, to
judge from numerous internal clues. Then, instead of publishing
the theory, he kept it to himself for almost fifty years!  This is
baffling, if we only ask ourselves how we might have acted.
Rather, we must ask how an intellectual of Galileo's time would
have acted, and why.

The lectures themselves were evidently a success, so what was the
crisis? Here we must note that the context of the lectures was a
dispute, a ``duel," between Manetti and Vellutello.  The lectures
are actually a savage attack on Vellutello.  To an extent that it
is hard to appreciate now, Renaissance intellectual life, at least
in the vicinity of a court, consisted in attacks and
counterattacks, which often functioned as a kind of spectator
sport \cite{Biagioli2}. This kind of ongoing dispute, with
reputations on the line, was the very stuff of Italian
intellectual life. Galileo's own career offers many examples. And
here was the crisis.  Nothing would be more natural than to
anticipate another round in this dispute, a reply to Galileo from
some partisan of Vellutello. Now an intellectual should expect to
be attacked -- it went without saying -- and he should have a good
reply ready.  A good reply could be something impressive, or
witty, or surprising -- any number of rhetorical strategies were
possible, but a good reply had to be tenable.  The worst disaster
was not to be wrong, but rather to have nothing adequate to say.
And Galileo's position on scaling in the lectures was untenable.
He had to develop a new position.

In this culture of attack and counterattack, it made a certain
kind of sense to keep results secret until they were needed. The
most famous example of this phenomenon is probably the solution of
the cubic equation, which was held secret for decades
\cite{Cardano} until it was finally published by Cardano in 1545.
In the same way, Galileo may have developed the scaling theory and
then postponed publishing it because the moment never actually
arrived when he needed it.

It is not pure conjecture that Galileo would have followed such a
strategy, keeping secret an important result until he needed it
for self-defense.  Indeed, in the controversy on floating bodies
we see him doing exactly this. This controversy began when Galileo
asserted, by Archimedes' Principle, that ice must be less dense
than water, since it floats.  His Aristotelian opponents asserted
that according to Aristotle ice, being cold, must be more dense
than water, and that ice floats, despite being more dense, because
it is broad and flat.  Galileo quickly disposed of this foolish
argument, but then his opponents found an example of something (a
chip of ebony) that really is denser than water and nonetheless
floats, if you carefully put it on the water's surface (because it
is broad and flat, they said).  At this point neither side in the
controversy had a satisfactory understanding of what they were
seeing.  Galileo's slightly odd response was to bring forward a
way to prove Archimedes' Principle by geometry, using some
mechanical ideas of balance that had long fascinated him. This was
clearly something he had kept ready for just such a moment, even
though it did not really address the problem of the floating ebony
chip.  He made some new observations of this anomalous floating
phenomenon (which we attribute mainly to surface tension) and
stretched his geometric theory to argue that Archimedes' Principle
could entirely account for it.  This strategy worked, in the sense
that he argued his opponents to a standstill.  He never convinced
them, but that was not the point. Rather, he had used his hitherto
secret knowledge in a successful defense.  As we noted earlier,
Galileo also used the scaling theory defensively in this
controversy, right at the end, as a clincher, so to speak
\cite{floatingparadox}. Thus it is not implausible that he had
kept it for just such a purpose.

When Galileo finally published the scaling theory, at the end of
his life, he hints at the story of its discovery, including his
original naivete about scale invariance. We have seen how Sagredo
endorses scale invariance at the beginning of {\bf Two New
Sciences}, but in fact all three participants do this.  On the
Second Day, Simplicio, who is often comic in just this way,
suddenly endorses scale invariance, as if he had just thought of
it, and Sagredo has to point out that he had said it the day
before, and this is what they have been talking about.  At this
point Salviati, who is Galileo's spokesman, says that he too had
assumed scale invariance once, until various observations showed
him the contrary \cite{TNS12}. In this way Galileo is commendably
frank, truthfully representing his former opinion in the Inferno
lecture. On the other hand, he never mentions the lecture itself,
which would still have been awkward for the honor of the
Florentine Academy. No one can blame him for this. Quite possibly
by 1638 there was no one else alive who even remembered this
detail of the original lecture. There was no real need to recount
this story. Thus the motivating problem of {\bf Two New Sciences},
the collapse of Manetti's Inferno, is discreetly avoided, and
another one, not particularly convincing, a ship that falls apart
of its own weight, is invented in its place.

In fact, Galileo seems to have regarded the Inferno lectures as an
embarrassment, and this may be another reason why he postponed
publishing the scaling theory.  His first biographer, Viviani, was
unaware of the lectures, despite living in Galileo's house during
his last years, collecting Galileo's stories, and devoting himself
to Galileo's memory after his death.  Since Viviani is the source
of most stories about Galileo, subsequent biographies have also
said little or nothing about the Inferno Lectures, which were only
rediscovered in the 19th century.  Thus one can plausibly conclude
that Galileo never told Viviani about them.  This is rather odd
when you think that they were very likely the pivotal opportunity
that got his career started. A letter of 1594 from one Luigi
Alamanni \cite{Alamanni} also indicates that Galileo was unhelpful
to someone who wanted to get a copy of them. The letter says
``About that lecture of Galileo, he is in Padua, and I have not
been able to get it from him." \cite{Alamanni2}  There were copies
with Bacio Valori in Florence, including one written out in
Galileo's own hand, but he apparently did not volunteer this.

There is another hint that Galileo's scaling theory was originally
planned defensively. In working out the scaling theory, Galileo
would have noticed a second problem in the Inferno lectures,
besides the inevitable collapse of the roof. In the first lecture
the dimensions of the lowest regions are determined by comparison
with certain giants who have been placed there by Dante, embedded
in ice. In the lecture Galileo assumed that these giants have the
same proportions as normal men, and his only hesitation on this
point was whether giants have the ideal human proportions favored
by artists like Albrecht D\"urer, who wrote on the subject, or
whether they have proportions more like ordinary men. Manetti and
Vellutello had differed on this point, and Galileo favored
Manetti, of course. But it is a consequence of the scaling theory
that giants couldn't have either of these proportions. If we look
in {\bf Two New Sciences} in the section on scaling in animals
\cite{TNS14}, we find that in fact the animal of most concern is
man: Galileo is concerned about {\em giant men}.  He even quotes a
poet describing a giant, although it is Ariosto, not Dante, and he
suggests, although it is a rather farfetched interpretation of the
words, that Ariosto understands that giants would be misshapen.
This occurrence of poetic giants and their proportions in {\bf Two
New Sciences} is entirely to be expected if the context is the
Inferno lectures. The Ariosto lines make sense as part of a
prepared defense: not only could he expound the scaling theory, he
could invoke Ariosto as a poet who understands it. Although this
might not impress a scientific audience, the audience that
mattered was a courtly audience.  And although they would not
understand the geometry, they would undoubtedly applaud this
trick, just as they had applauded his skillful combination of
Dante and geometry in the original lecture. Nearly fifty years
later, as he finally wrote the scaling theory down, for a very
different audience, these associations still persist. This is
speculation, of course, but the reader is invited to find a more
plausible explanation for what Ariosto's giant is doing in {\bf
Two New Sciences}.

The setting in the Venetian arsenal seems to promise practical
science, and the scene is so effective that it has functioned as a
credential for Galileo ever since, showing him to be a practical
man of science, right at home with the foreman on the shop floor.
There can be no doubt whatever that Galileo really was an
ingenious and skilful experimentalist, but this opening scene
fails to suggest how theoretical the work on scaling actually is.
The Arsenal fades away almost immediately on the First Day, never
to reappear. The discussion on the Second Day, when the subject
finally comes into focus, is about beams with rectangular and
circular cross section because the aim is geometrical proof. And
the result toward which the whole development is tending is
entirely theoretical, and is given no practical illustration,
namely that {\em anything} would break of its own weight if it
were sufficiently scaled up.  This is the content of the
culminating Propositions VII and VIII on the Second Day.
Proposition VII, for example, is ``Among heavy prisms and
cylinders of similar figure, there is one and only one which under
the stress of its own weight lies just on the limit between
breaking and not breaking: so that every larger one is unable to
carry the load of its own weight and breaks; while every smaller
one is able to withstand some additional force tending to break
it."  This 17th century version of ``the inevitability of
gravitational collapse" is what the Second Day is really about.

It is noteworthy that Galileo describes no experiments to verify
the scaling theory, although experiments on the bending of a
loaded beam would have been very simple for him, and entirely
characteristic of his Padua years.  This circumstance suggests
that the scaling theory is early, and had no practical motivation.

The Inferno lectures begin by praising the skill and audacity of
those discoverers who have measured the heavens, and the surface
of the earth, and point out how much more difficult it is to know
the earth's interior, where it seems that no one can go and return
-- yet even this our Dante has done! It is an implied theme of the
lectures that the geometer can do this too:  by the use of
geometry we see beyond the limitations of our senses, reasoning
about otherwise inaccessible things. This insight is, in a way,
the heart of physics as Galileo came to understand it. Yet in the
Inferno lectures the idea occurs almost accidentally, as part of a
glib and clever entertainment. The geometry is merely descriptive.
There are no actual discoveries.

It is entirely different with the scaling theory, motivated by the
lectures.  Suddenly geometry really {\em did} give the key to new
discovery, a discovery that Galileo was still proud to claim as
his own at the end of his amazing career.  This must have come as
a wonderful surprise. For all the faith that Galileo and a few
other mathematical enthusiasts had in geometry, it was, in
practice, the everyday tool of artisans and merchants, not the
source of new insights.  Archimedes, to be sure, had used geometry
in this wonderful way, but no one imagined there could be another
Archimedes.  With the scaling theory, however, Galileo had
something truly new, worthy of comparison with Archimedes,
something that validated his faith in geometry and hinted at
undreamed of successes to come.

Looked at this way, Galileo's lifelong reluctance to publish seems
even more inexplicable, but perhaps this pattern began with the
traumatic experience of the Inferno Lectures. He seems to have
done his best to make people forget the lectures, and he kept the
scaling theory to himself. What he made public, at least in this
case, was a source of trouble, while what he kept secret was a
source of confidence. This is a lesson that he might have taken to
heart then, and it is a view he explicitly voices later on in the
opening lines of {\bf The Assayer}. Galileo frequently claims to
have wonderful results that he has not yet revealed. We know that
this was true through much of his career, and apparently it was
true right from the start.

Finally it seems a fine irony that the first success of Galileo's
mathematical physics, which is close to being the first success of
mathematical physics at all, was a response to a problem that was
not physical, but rather the collapse of an imaginary structure in
a work of literature.
\section*{Acknowledgements}
I would like to thank two anonymous referees for helpful
suggestions and criticisms.


\begin{thebibliography}{***}
\bibitem{conformal} The same aesthetic idea is the
motivation for conformal field theories.
\bibitem{TNS7} Archimedes' law of the lever, the balance of
torques, was second nature to Galileo.
\bibitem{secondmoment} We would now say the stress itself varies
linearly across the beam, so that the correct dependence is the
second moment of area, going as the diameter to the fourth power.
\bibitem{TNS6} All propositions in this paragraph are from the
second day.
\bibitem{Manetti}  In fact Manetti was a Florentine intellectual
of the previous century, the biographer of Filippo Brunelleschi.
He lived well before the founding of the Florentine Academy.
\bibitem{Vellutello} Vellutello was from Lucca, a notable rival
of Florence.
\bibitem{Alamanni2} Since he couldn't obtain the text, Alamanni
provided his correspondent a summary of the lectures (five years
after they were delivered!):  ``It consisted in this, that he
reviewed the opinion of the Florentine Antonio Manetti concerning
the site of Dante's Inferno, published in a book printed by the
Giunti, and then he reviewed the opinion of Vellutello, a
commentator on Dante, on the same subject, and comparing the one
with the other, he showed that of Manetti to be better."
\end{thebibliography}
\end{document}